\documentclass{appolb}
\usepackage{graphicx}
\usepackage{physics}
\usepackage{amsmath}
\usepackage{amssymb}
\usepackage{siunitx}
\usepackage[backend=biber]{biblatex}
\usepackage{booktabs}
\usepackage{multirow}
\usepackage{tikz}
\usepackage{tikz-feynman}
\usetikzlibrary{matrix, arrows.meta, positioning}
\begin{document}
% \vspace{-4cm
\title{\textsc{BabaYaga@NLO} at present and future $e^+e^-$ colliders
\thanks{Presented at Matter To The Deepest Recent Developments In Physics Of Fundamental Interactions XLVI International Conference of Theoretical Physics, 14-19 September 2025, Katowice, Poland.}\\
\small Celebrating 25 years of BabaYaga
% you can use '\\' to break lines
}
\author{
Francesco P. Ucci
\address{Dipartimento di Fisica "Alessandro Volta", Università di Pavia, Via A. Bassi 6, 27100 Pavia, Italy\\
INFN, Sezione di Pavia, Via A. Bassi 6, 27100 Pavia, Italy}
}
\maketitle
\begin{abstract}
Precise QED radiative corrections for low- and high-energy electron-positron colliders are essential for accurate simulations of luminosity processes and precision tests of the Standard Model. We review the historical formulation and the recent developments of the \textsc{BabaYaga@NLO} event generator, which implements a QED Parton Shower matched with fixed-order calculations. 
We discuss the theoretical formulation of the code, as well as the assessment of its theoretical accuracy. Applications at low- and high-energy $e^+e^-$ colliders are presented, including latest result, together with the perspectives for future improvements, in view of the demanding precision requirements of future machines at the intensity frontier. 
\end{abstract}
  
\section{Introduction}
Precision measurements at electron-positron colliders rely on the accurate theoretical description of Quantum Electrodynamics (QED) radiative corrections. Photon radiation from the beams and charged final states can affect cross section and event shapes~\cite{Frixione:2022ofv}, therefore its precise modelling is mandatory to reach the sub-percent accuracy required by Standard Model (SM) precision measurements.
An example of the importance of QED radiative corrections and PS algorithms is given by luminosity measurements~\cite{CarloniCalame:2015zev}, for which a precision below $0.1\%$ ($0.01\%$) leven is required at low(high)-energy machines. A comparable level of precision is also required for the measurement of $e^+e^-\to\textit{hadrons}$ cross section at centre of mass (c.m.) energies below $2~\rm{GeV}$~\cite{WorkingGrouponRadiativeCorrections:2010bjp}. This quantity enters the dispersive calculation of the anomalous magnetic moment of the muon $a_\mu$ in the Standard Model, which exhibits a several-$\sigma$ discrepancy with data~\cite{Aliberti:2025beg}. \\
For instance, on the theory side, the relevant logarithm appearing in Bhabha scattering luminosity -- $L\equiv \log(Q^2/m_e^2)$ where $Q^2$ is the relevant energy scale squared and $m_e$ the electron mass -- is of the order of $15-20$ at the energy scale of $1-10~\rm{GeV}$, considering a large angular acceptance at flavour factories, or a small one at high-energy machines. For these reasons, an accurate description of additional photon radiation is of utmost importance, making Monte Carlo generators for QED radiative corrections an essential tool for $e^+e^-$ colliders phenomenology. A description and comparison between the available codes is available in~\cite{Aliberti:2024fpq}.\\
 In this context, Parton Shower (PS) algorithms play a central role, by providing an event-by-event description of photon emissions in an exclusive way, therefore allowing for the resummation of large logarithms appearing in perturbation theory -- see Tab.~\ref{tab:logstructure} -- while preserving the differential structure of the final states. This latter feature is particularly important for realistic MC event generation, where all relevant experimental event selection criteria and acceptance cuts can be implemented.\\
In this work, we will review the theoretical description of the \textsc{BabaYaga} Parton Shower algorithm, retracing the main steps that brought to the actual formulation of \textsc{BabaYaga@NLO}, as well as some applications of the code in relevant experimental setups, where it has been used during the 25 years of its development. 
\begin{table}[t]
\centering
\renewcommand{\arraystretch}{1.4}
\begin{tabular}{r| l l l}
% \hline
 & \textbf{LL} & \textbf{NLL} & \textbf{NNLL} \\
\hline
\textbf{LO}   
& $\alpha^0$ 
&  
&  \\
\textbf{NLO}  
& $\alpha L$ 
& $\alpha$ 
&  \\
\textbf{NNLO} 
& $\tfrac{1}{2}\alpha^2 L^2$ 
& $\tfrac{1}{2}\alpha^2 L$ 
& $\tfrac{1}{2}\alpha^2$ \\
\textbf{H.O.} 
& $ \sum_{n=3}^{\infty}\frac{1}{n!}\alpha^n L^n$ 
& $ \sum_{n=3}^{\infty}\frac{1}{n!}\alpha^n L^{n-1}$ 
& $\dots$ \\
% \hline
\end{tabular}
\caption{Perturbative structure of QED radiative corrections. From top to bottom, the power countingin $\alpha$ increases the fixed-order accuracy. From right to left, the logarithmic expansion $L\equiv \log{Q^2/m_e^2}$ becomes more relevant.}
\label{tab:logstructure}
\end{table}
\section{The QED Parton Shower algorithm}
The original version of \textsc{BabaYaga}~\cite{CarloniCalame:2000pz} was developed for luminosity measurements at flavour factories with 
Large angle Bhabha scattering (LABH), within the leading-logarithmic (LL) approximation; then, an updated version~\cite{Balossini:2006wc} introduced the Matching with next-to-leading (NLO) exact matrix elements, reducing the theoretical error to the $0.1\%$ level. In this Section, we review the main ideas underlying the theoretical formulation of the QED Parton Shower algorithm implemented in the most recent version \textsc{BabaYaga@NLO} (see~\cite{CarloniCalame:2001ny} and references therein for a detailed review).
\subsection{A Monte Carlo solution of DGLAP equations}
The cross section for the generic process $e^+e^-\to X^+X^-$ at c.m. energy $\sqrt{s}$, including the effect of photon radiation emitted from charged legs, can be written as~\cite{PhysRevD.48.1021}
\begin{equation}
    \sigma(s)=\prod_{i}\int \dd x_{i}D\left(x_i,Q^2\right)\int \dd\Omega \dv{\sigma_0}{\Omega}{} (x_1x_2s)\Theta(\text{cuts})
\end{equation}
where $i=1,2\, (3,4)$ is the energy fraction of the initial (final) state particles entering the hard-scattering process described by the Born-level differential cross section $\dd \sigma_0$, while the Heaviside $\Theta$ implements the relevant cuts on the phase-space $\dd \Omega$. The Structure Functions (SF) $D(x,Q^2)$ are an effective way to describe multiple emissions of hard and soft photons collinear to the emitting particles, and are solution of the Dokshitzer-Gribov-Lipatov-Altarelli-Parisi (DGLAP) equation~\cite{DGLAP}.
 The SF con be written as the sum of terms each expressing the $n$-th photon emission~\cite{CarloniCalame:2001ny}
\begin{equation}
\renewcommand{\arraystretch}{2.2} % Aumentato leggermente per dare spazio al diagramma
\begin{array}{l l}
    % --- Prima riga con diagramma ---
    D(x,s) = &\vspace{-0.5cm}\\
    \begin{tikzpicture}[baseline=-0.6ex] % Allinea il centro del diagramma al testo
        \begin{feynman}
            \vertex (a);
            \vertex (b) at (2cm, 0); % Forza la lunghezza a 2cm esatti
            \diagram* {
                (a) --[fermion] (b),
            };
        \end{feynman}
    \end{tikzpicture}
     % Spazietto estetico
   & \quad \Pi(s,m^2)\,\delta(1-x)
    \\
      \begin{tikzpicture} % Allinea il centro del diagramma al testo
        \begin{feynman}
            \vertex (a);
            \vertex (b) at (2cm, 0); %
            \vertex(c) at (0.5cm,0cm);
            \vertex(d) at (1cm,0.5cm);
            \diagram* {
                (a) --[fermion] (b),
                (c)--[photon](d),
            };
        \end{feynman}
    \end{tikzpicture}
    % --- Seconda riga (Integrale) ---
    &
    +\displaystyle
    \frac{\alpha}{2\pi}
    \int_{m^2}^{s} \frac{\mathrm{d}s'}{s'}\,
    \Pi(s,s')\,\Pi(s',m^2)
    \int_0^{1-\varepsilon} \mathrm{d}y\,
    P(y)\,\delta(x-y)
    \\
\begin{tikzpicture} % Allinea il centro del diagramma al testo
        \begin{feynman}
            \vertex (a);
            \vertex (b) at (2cm, 0); %
            \vertex(c) at (0.5cm,0cm);
            \vertex(d) at (1cm,0.5cm);
             \vertex(e) at (1.5cm,0cm);
            \vertex(f) at (2cm,0.5cm);
            \diagram* {
                (a) --[fermion] (b),
                (c)--[photon](d),
                (e)--[photon] (f),
            };
        \end{feynman}
    \end{tikzpicture}
    % --- Seconda riga (Integr
    % --- Terza riga (2 fotoni) ---
    &
    + \text{2 photon terms...}
\end{array}\label{eq:DxsPS}
\end{equation}
by introducing the Sudakov form factor $\Pi(s,s')$, which represents the probability that the parton $X$ evolves from the virtuality $s'$ to $s$ emitting a photon with energy fraction below $\varepsilon$.
The MC implementation is based on updating the virtualities of each parton following Eq.~\eqref{eq:DxsPS}, as detailed in~\cite{CarloniCalame:2000pz}, therefore generating the energy of each photon accordingly. The PS approach has the advantage of allowing for an exclusive generation of the photon angular spectrum beyond the collinear approximation. In \textsc{BabaYaga}, inspired by the YFS approach~\cite{YFS1961}, the generation of a transverse momentum $k_{l,T}\neq 0$ of the $l$-th photon is achieved by generating the $\cos\theta_l$ as the eikonal function $\mathcal{I}_{ij}(k_l)$, calculated with respect to the external charged legs $i,j$.
%\begin{equation}
% \mathcal{I}_{ij}(k_l)= \eta_i\eta_j \frac{\left(p_i\cdot p_j\right)}{\left(p_i\cdot k_l\right)\left(p_j\cdot k_l\right)}(k^0_l)^{2}\,.
%   \label{eq:eik}
% \end{equation}
By combining all these ingredients, we arrive at the expression of the cross section in PS approach at leading-logarithmic (LL) accuracy
\begin{equation}
    {\rm d}\sigma_{\rm PS}^{\rm LL }=\Pi (\varepsilon \sqrt{s}/2,Q^2)\sum_{n=0}^\infty \frac{1}{n!} \left|\mathcal{M}_n^{\rm LL}\right|^2 {\rm d}\Phi_n 
    \label{eq:PSLL}
\end{equation}
where the Sudakov form factor is defined as
\begin{equation}
    \Pi(\varepsilon,Q^2)=\exp\left\{-\frac{\alpha}{2\pi}\int_0^{1-\varepsilon}{\dd}z\, P(z) \int {\dd} \Omega_kI(k)\right\}\,,
\end{equation}
and $P(z)$ is the Altarelli-Parisi vertex for the $X\to X+\gamma$ branching, which describes the probability density of emitting a photon with energy fraction $1-z$ of the parent parton $X$ and with $ I(k)=\sum_{ij}\mathcal{I}_{ij}(k)$.
The matrix element for the $n$-th photon emission $\mathcal{M}_n^\text{LL}$ is an approximation of the exact $\mathcal{M}_n$ at LL accuracy, where each photon emission is approximated by a factor $\alpha/(2\pi) P(z)I(k)$ calculated on a kinematics mapped onto the $2\to2$ underlying Born process and factorised on $|\mathcal{M}_0|^2$, that can be iterated from the $n=1$ case. \\
\subsection{BabaYaga@NLO: matching and precision}
In the LL approximation, all the tower of logarithms $\alpha^n L^n$, corresponding to the first column of Tab.~\ref{tab:logstructure}, is correctly resummed, with the $\order{\alpha}$ being the first missing term.
Introduced in~\cite{Balossini:2006wc}, the matching procedure allows us to correctly include the exact NLO calculation while resumming all the $n\geq 2$ photon emissions. The procedure can be easily understood by comparing the $\order{\alpha}$ expansion of Eq.~\eqref{eq:PSLL} with the exact calculation. Therefore, one can introduce the \textit{soft-virtual} correcting factor in a multiplicative way $F_\text{SV}=1-\left(C_\alpha-C_\alpha^\text{LL}\right)$ -- where $C_\alpha^{(\text{LL})}$ is the coefficient multiplying the zero-photon exact (LL) soft+virtual cross section -- 
so that the exponentiated $\order{\alpha}$ correction is always exact. In the same spirit, the real matrix elements can be corrected by introducing an \textit{hard} correction
$   F_\text{H}=1+(|\mathcal{M}_1|^2-|\mathcal{M}_1^\text{LL}|^2)/|\mathcal{M}_1|^2
$
arriving at the \textsc{BabaYaga@NLO} master formula, valid for any number of photonic emissions
\begin{equation}
\dd\sigma_\text{NLOPS}=F_\text{SV}\,\Pi(\varepsilon\sqrt{s}/2,\{p\})\sum_{n=0}^\infty\frac{1}{n!}\left(\prod_{i=1}^n F_{\text{H},i}\right)|\mathcal{M}_n^\text{LL}|^2\dd\Phi_n(\{p\},\{k\})\,.
\end{equation}
The first term that we are missing in this approximation is the $\order{\alpha^2 L}$, which is $(\frac{\alpha}{\pi})^2L\sim 0.1\%$ at $Q^2\sim \order{\rm{GeV}}$, giving a rough estimate of the theoretical precision of the approach. By a tuned comparison with other available codes, such as \textsc{BHWIDE}~\cite{Balossini:2006wc,},  as well with analytical calculations~\cite{CarloniCalame:2011zq}, the accuracy of \textsc{BabaYaga@NLO} is assessed at the $0.1\%$ level at flavour factories, as it will be discussed later on.
% \subsection{Theoretical precision}
\subsection{Available final states}
To conclude this Section, we briefly show in Tab~\ref{tab:babayaga_processes} all the available processes in the current and upcoming version of \textsc{BabaYaga@NLO}, with the associated precision at which are calculated. 
\begin{table}[h]
\centering
\renewcommand{\arraystretch}{1.3}
\begin{tabular}{l l l l l}
\toprule
 & \textbf{Process} & \textbf{Order (NLOPS QED $\mathbf\oplus$)}& \textbf{Ref.} \\
\midrule
\multirow{4}{*}{\rotatebox{90}{\textbf{Published}}}
& $e^+e^-$ 
&  LO EW $\oplus$ LO SMEFT
& \cite{CarloniCalame:2000pz,Balossini:2006wc}
\\
& $\mu^+\mu^-$
&  LO EW
& \cite{Balossini:2006wc}
\\
& $\gamma\gamma$
&  NLO EW
& \cite{CarloniCalame:2011zq}
\\
& $\pi^+\pi^-$
&  FF (F$\times$sQED, GVMD, FsQED)
& \cite{Budassi:2024whw}
\\
\midrule
\multirow{3}{*}{\rotatebox{90}{\textbf{WIP}}}
& $e^+e^-\gamma$
&
& \multirow{3}{*}{\cite{Budassi:2026}}
\\
& $\mu^+\mu^-\gamma$
&
&
\\
& $\pi^+\pi^-\gamma$
& FF (F$\times$sQED)
& 
\\
\bottomrule
\end{tabular}
\caption{Processes ($e^+e^-\to XX$) implemented in \textsc{BabaYaga@NLO} and corresponding perturbative accuracy, where the NLOPS QED is intended for all final states. FF stands for Form Factor, included in the NLO calculation for $\pi\pi$ final states, with the approach indicated in parenthesis.}
\label{tab:babayaga_processes}
\end{table}\\
In the following Sections, we show some phenomenological applications of radiative corrections, as computed with \textsc{Babayaga}, in relevant experimental scenarios.
\section{Low-energy colliders}
Electron-positron colliders operating in the $1-10~\rm{GeV}$ energy range continue to play a central role in precision tests of the Standard Model and in hadronic physics. Experiments such as KLOE, BESIII, BelleII, BaBar, CMD, SND are either still taking data or analysing high-statistics datasets, which require a very good control of QED radiative corrections. One of the primary goals of these machines is the precise measurement of hadronic final states, which enter the dispersive calculation of the leading-order hadronic vacuum polarisation contribution to the muon $g-2$ anomaly,
\begin{equation}
a_\mu^{\text{HVP},\text{LO}}=\left(\frac{\alpha m_\mu}{3\pi}\right)^2\int_{m_{\pi_0}^2}^\infty\dd s\frac{\hat{K}(s)}{s^2}R(s)
\label{Eq:disprel}
\end{equation}
where $R(s)=\sigma(e^+e^-\to\textit{hadrons}\,(+\gamma))/\sigma_0(e^+e^-\to\mu^+\mu^-)$ is the ratio between the photon-inclusive hadronic cross section to the tree-level dimuon one. In this framework, Monte Carlo generators fulfil various tasks: they are employed for precision luminosity measurements via Large Angle Bhabha Scattering (LABS) or di-photon production, for QED tests through the measurement of the ratios like $N_{e^+e^-}/N_{\mu^+\mu^-}$, and for the modelling of exclusive hadronic channels such as $\pi^+\pi^-,K_LK_S,\dots$.
In this Section, we present two examples to illustrate the crucial role played by Monte Carlo generators in low-energy precision measurements.
\subsection{Luminosity measurements at flavour factory}
In collider physics, the luminosity $\mathcal{L}$ is a fundamental machine parameter that allows to convert the observed number of events for a certain process to its absolute cross section, via the relation $\sigma = N_\text{exp}/\mathcal{L}$. At $e^+e^-$ machines, it is convenient to measure the luminosity using a well-defined reference process, exploiting the relation $\mathcal{L}=N_0/\sigma_0^\text{th}$, whose cross section $\sigma_0$ is calculable with very high accuracy in perturbation theory. At flavour factories, luminosity is tipically  measured with a relative error of some per-mille, making the resummation of multiple photon emissions a mandatory requirement for MC generators. \\
In experimental analyses, the theoretical precision is often estimated by comparing two independent generators. \textsc{BabaYaga3.5}~\cite{CarloniCalame:2003yt} was used as a benchmark in a number of experiments, with a theoretical precision of $0.5\%$ for Large-angle Bhabha and of $1\%$ for di-photon and $\mu^+\mu^+$ production, due mainly to the missing $\order{\alpha}$ constants~\cite{WorkingGrouponRadiativeCorrections:2010bjp}. The matched version \textsc{BabaYaga@NLO} has improved the theoretical precision to the level of $0.1\%$, as shown by tuned comparison~\cite{WorkingGrouponRadiativeCorrections:2010bjp} for the Bhabha channel at $\phi,\tau,B$ factories, as well as for the $\mu^+\mu^-$~\cite{Aliberti:2024fpq} in a CMD-like scenario.
\subsection{The Pion form factor}
The cross section of the process $e^+e^-\to\pi^+\pi^-$, parametrised with the Pion form factor $F_\pi(q^2)$, is the dominant contribution to $a_\mu^{\text{HVP},\text{LO}}$, due to the peaking behaviour of the dispersion integral~\eqref{Eq:disprel} in the low-energy region. It can be alternatively extracted from energy-scan measurements or by the radiative return through the $e^+e^-\to\pi^+\pi^-\gamma$ process, with current experimental determinations exhibiting tensions up to $5\sigma$ level~\cite{Aliberti:2025beg} between the two methods. Understanding the origin of this discrepancy needs a very good control of radiative corrections and, in particular, the availability of independent MC generators for the $\pi^+\pi^-(\gamma)$ final state. 
For example, in the latest measurement of $F_\pi(q^2)$ by the CMD-3 collaboration~\cite{CMD-3:2023alj} via energy scan, the uncertainty associated to radiative corrections is the $0.3\%$ -- more than $40\%$ of the systematic error -- mostly coming from the differences between \textsc{BabaYaga@NLO} and \textsc{MCGPJ}. In this context, the latest developments of \textsc{BabaYaga@NLO}~\cite{Budassi:2024whw} are in the direction of improving the theoretical description of the $\pi\pi$ channel, by implementing PS algorithm, taking into account the internal structure of the pion at NLO accuracy. In particular, beyond the traditional scalar QED approach (F$\times$sQED), the form factor has been introduced in the Generalised Vector-Meson Dominance (GVMD) model~\cite{Ignatov:2022iou} and in the dispersive approach (FsQED)~\cite{Colangelo:2022lzg}. These improved descriptions of $\pi^+\pi^-$ show a very good agreement with the charge asymmetry data, an observable extremely sensitive to the treatment of $F_\pi(q^2)$ in the loops. Ongoing efforts~\cite{Budassi:2026} aim to generalise the formulation of \textsc{BabaYaga@NLO} also for $\mu^+\mu^-\gamma, \,\,\pi^+\pi^-\gamma$ final states, representing the first calculation at NLOPS accuracy for radiative processes.
\section{Future high-energy colliders}
Next-generation electron-positron colliders are being designed as Higgs, Top, and Electroweak (EW) factories~\cite{Altmann:2025feg}, allowing for precision measurements at energies ranging from the $Z$-pole up to the $t\bar t$ production threshold. Luminosity calibration will play a crucial role in revisiting the Electroweak measurements, in order to fully exploit the very high statistics expected. We focus the discussion on the FCC-ee~\cite{FCC:2025lpp}, which is expected to operate at c.m. energies of $
    \sqrt{s}= 91,\, 160,\, 240,\, 365 \,\rm{GeV}.
$
Precision measurements at FCC-ee, such as the $W$ boson mass and width, will require a luminosity uncertainty at the level of $\delta\mathcal{L}\leq 10^{-4}$, while for the measurement of the $HZ$ production cross section $\delta \mathcal{L}$ should be kept at least at the per-mille level. Building on the experience of LEP, luminosity at future machines is expected to be measured either with SABS or $e^+e^-\to\gamma\gamma$ at large angle. In this Section we presenting some recent developments in \textsc{BabaYaga} at high energy, including NLO electroweak corrections to the $\gamma\gamma$ and potential New Physics (NP) contamination in SABS.
\subsection{High precision $\gamma\gamma$ luminosity measurements}
The diphoton production, a pure QED process at leading order, is already used as for luminosity calibration at few-$\rm{GeV}$ flavour factories with per-mille accuracy~\cite{Balossini:2008xr}. At high energy, $\gamma\gamma$ events are in principle statistically more limited w.r.t. the large-angle Bhabha -- $\sigma_\text{LABS}\simeq 66\, \sigma_{\gamma\gamma}$ at $\sqrt{s}=91~\rm{GeV}$ -- which acts as a background. On the other hand, the $e^+e^-\to\gamma\gamma$ cross section has the advantage of a reduced sensitivity to the HVP, which enters only at two-loop level, making it a theoretically sound alternative to the Bhabha. Moreover, experimentally, reaching 10ppm ultimate relative precision on SABS events is technically challenging~\cite{Altmann:2025feg}, requiring to control the position of calorimeters at sub-$\mu \textrm{m}$ level. \\
In order to meet the $10^{-4}$ precision goal, at least NLO electroweak corrections have to be included with higher orders QED on top. This has been done in~\cite{CarloniCalame:2019dom,Carloni:2019ejp}, where NLO EW corrections have been estimated to be $\order{1-2\%}$ in angular distribution, in a $\ang{20}-\ang{160}$ setup, of the same size of QED higher orders. Moreover, the effect of the fermionic and hadronic NNLO corrections, with their uncertainty, has been estimated by factorising the vacuum polarisation on the pure NLO corrections as
\begin{equation}
\sigma_\text{NNLO}^{\Delta\alpha}
\pm \delta\sigma_\text{had}\simeq (\sigma_\text{NLO}^
\text{QED}-\sigma_\text{LO}) \times\left[\Delta \alpha (s)\pm\Delta\alpha_\text{had}\right]\,
\end{equation}
being of the order or $0.1\%$ at FCC-ee. In order to lower the uncertainty to the $0.01\%$ level, a first step would be the inclusion of the exact $\order{\alpha^2 N_f}$ fermionic virtual and real corrections, which is in progress in \textsc{BabaYaga}, as well as the full NNLO QED corrections, which have been implemented recently in \textsc{McMule}~\cite{Engel:2025fmp}, possibly matched with a PS.
\subsection{New physics in SABS luminosity}
On top of radiative corrections, one may ask if any unknown New Physics could contaminate luminosity processes at a level comparable with the FCC-ee precision target. A first step in this direction has been made in~\cite{maestre2022precisionstudiesquantumelectrodynamics} for $e^+e^-\to\gamma\gamma$, while the latest update of \textsc{BabaYaga} concerns NP in SABS. In the Standard Model, the Bhabha at small angle is largely dominated by the $t$-channel photon exchange, being almost a pure QED process. At $10^{-4}$ level, on top of NLOPS EW corrections, the process has an uncertainty due to the HVP. In the hypothesis of such error to be reduced by the time of future colliders, we investigated whether any light or heavy New Physics could contaminate the process at the precision goal. Based on present bounds on beyond the SM interactions, it has been found that Light NP has negligible effects, while the Heavy NP, parameterised by means of the Standard Model Effective Field Theory (SMEFT) has an effect in the range of $10^{-5}-10^{-4}$, representing a potential source of uncertainty for FCC-ee at the $Z$ pole. A strategy to reduce such contaminations could be to measure the forward-backward asymmetry for the LABS, providing therefore model-independent constraints on Heavy NP by fitting $A_{\text {FB}}(\sqrt{s})$.
\section{Outlook and future developments}
QED radiative corrections are a fundamental ingredient for low- and high-energy electron-positron colliders, where the resummation of multiple photon emission is mandatory for precision studies. The \textsc{BabaYaga} event generator has been developed for precision luminosity measurements at flavour factories more than 25 years ago, and has received continuous improvement. The latest version \textsc{BabaYaga@NLO} is able to generate many processes at NLOPS in QED, with an estimated theoretical accuracy of $0.1\%$. In this work, we have shown the theoretical formulation of the code, as well as some of the historical and more recent phenomenological applications. In the next future, an important update regarding radiative processes $X^+X^-\gamma$, for $X=\mu,\pi$ will be released, marking a milestone for the measurement of hadronic cross sections at low energy.\\
In the future, the \textsc{BabaYaga} team plans to further improve the code by introducing the next-to-next-to-leading order matching with the Parton Shower, proceeding on the path to $0.01\%$ precision for future $e^+e^-$ colliders.
\section*{Acknowledgements}
The author is indebted to C. M. Carloni Calame, G. Montagna, O. Nicrosini, and F. Piccinini, the original authors of \textsc{BabaYaga}. We also thank E. Budassi, M. Ghilardi, A. Gurgone, and M. Moretti for fruitful collaboration. The author is also grateful to the Instituto de Física Teórica UAM-CSIC for its hospitality.
%uncomment the following lines to place a figure
%\begin{figure}[htb]
%\centerline{%
%\includegraphics[width=12.5cm]{Fig1}}
%\caption{Plot of ...}
%\label{Fig:F2H}
%\end{figure}
\printbibliography

@inproceedings{Frixione:2022ofv,
    author = "Frixione, S. and others",
    title = "{Initial state QED radiation aspects for future $e^+e^-$ colliders}",
    booktitle = "{Snowmass 2021}",
    eprint = "2203.12557",
    archivePrefix = "arXiv",
    primaryClass = "hep-ph",
    reportNumber = "FERMILAB-PUB-22-129-T",
    month = "3",
    year = "2022"
}

@article{CarloniCalame:2015zev,
    author = "Carloni Calame, C. M. and Montagna, G. and Nicrosini, O. and Piccinini, F.",
    title = "{High$-$precision Luminosity at $e^+ e^-$ Colliders: Theory Status and Challenges}",
    doi = "10.5506/APhysPolB.46.2227",
    journal = "Acta Phys. Polon. B",
    volume = "46",
    number = "11",
    pages = "2227",
    year = "2015"
}

@article{Aliberti:2024fpq,
    author = "Aliberti, Riccardo and others",
    title = "{Radiative corrections and Monte Carlo tools for low-energy hadronic cross sections in $e^+ e^-$ collisions}",
    eprint = "2410.22882",
    archivePrefix = "arXiv",
    primaryClass = "hep-ph",
    doi = "10.21468/SciPostPhysCommRep.9",
    month = "10",
    year = "2024"
}

@article{Aliberti:2025beg,
    author = "Aliberti, R. and others",
    title = "{The anomalous magnetic moment of the muon in the Standard Model: an update}",
    eprint = "2505.21476",
    archivePrefix = "arXiv",
    primaryClass = "hep-ph",
    reportNumber = "CERN-TH-2025-101, FERMILAB-PUB-25-0344-T, INT-PUB-25-015, IPARCOS-UCM-25-029, KEK Preprint 2025-22, LTH 1403, MITP-25-037, UWThPh 2025-15, UWThPh
  2025-15, ZU-TH 37/25, IPARCOS-UCM-25-029",
    doi = "10.1016/j.physrep.2025.08.002",
    journal = "Phys. Rept.",
    volume = "1143",
    pages = "1--158",
    year = "2025"
}

@article{CarloniCalame:2001ny,
    author = "Carloni Calame, Carlo Michel",
    title = "{An Improved parton shower algorithm in QED}",
    eprint = "hep-ph/0103117",
    archivePrefix = "arXiv",
    reportNumber = "FNT-T-2001-06",
    doi = "10.1016/S0370-2693(01)01108-X",
    journal = "Phys. Lett. B",
    volume = "520",
    pages = "16--24",
    year = "2001"
}

@article{WorkingGrouponRadiativeCorrections:2010bjp,
    author = "Actis, S. and others",
    collaboration = "Working Group on Radiative Corrections, Monte Carlo Generators for Low Energies",
    title = "{Quest for precision in hadronic cross sections at low energy: Monte Carlo tools vs. experimental data}",
    eprint = "0912.0749",
    archivePrefix = "arXiv",
    primaryClass = "hep-ph",
    reportNumber = "BIHEP-TH-2009-005, BU-HEPP-09-08, CERN-PH-TH-2009-201, FNT-T-2009-03, FREIBURG-PHENO-09-07, HEPTOOLS-09-018, IEKP-KA-2009-33, LNF-09-14-P, LPSC-09-157, LPT-ORSAY-09-95, LTH-851, MZ-TH-09-38, PITHA-09-14",
    doi = "10.1140/epjc/s10052-010-1251-4",
    journal = "Eur. Phys. J. C",
    volume = "66",
    pages = "585--686",
    year = "2010"
}

@article{CarloniCalame:2000pz,
    author = "Carloni Calame, C. M. and Lunardini, C. and Montagna, G. and Nicrosini, O. and Piccinini, F.",
    title = "{Large angle Bhabha scattering and luminosity at flavor factories}",
    eprint = "hep-ph/0003268",
    archivePrefix = "arXiv",
    reportNumber = "FNT-T-2000-05, SISSA-28-2000-EP",
    doi = "10.1016/S0550-3213(00)00356-4",
    journal = "Nucl. Phys. B",
    volume = "584",
    pages = "459--479",
    year = "2000"
}

@article{Balossini:2006wc,
    author = "Balossini, Giovanni and Carloni Calame, Carlo M. and Montagna, Guido and Nicrosini, Oreste and Piccinini, Fulvio",
    title = "{Matching perturbative and parton shower corrections to Bhabha process at flavour factories}",
    eprint = "hep-ph/0607181",
    archivePrefix = "arXiv",
    reportNumber = "FNT-T-2006-05",
    doi = "10.1016/j.nuclphysb.2006.09.022",
    journal = "Nucl. Phys. B",
    volume = "758",
    pages = "227--253",
    year = "2006"
}

@article{Balossini:2008xr,
    author = "Balossini, G. and Bignamini, C. and Calame, C. M. Carloni and Montagna, G. and Nicrosini, O. and Piccinini, F.",
    title = "{Photon pair production at flavour factories with per mille accuracy}",
    eprint = "0801.3360",
    archivePrefix = "arXiv",
    primaryClass = "hep-ph",
    reportNumber = "FNT-T-2008-01, SHEP-08-05",
    doi = "10.1016/j.physletb.2008.04.007",
    journal = "Phys. Lett. B",
    volume = "663",
    pages = "209--213",
    year = "2008"
}

@article{CarloniCalame:2011zq,
    author = "Carloni Calame, C. and Czyz, H. and Gluza, J. and Gunia, M. and Montagna, G. and Nicrosini, O. and Piccinini, F. and Riemann, T. and Worek, M.",
    title = "{NNLO leptonic and hadronic corrections to Bhabha scattering and luminosity monitoring at meson factories}",
    eprint = "1106.3178",
    archivePrefix = "arXiv",
    primaryClass = "hep-ph",
    reportNumber = "DESY-11-080, FNT-T-2011-01, LPN-11-25, SFB-CPP-11-26, WUB-11-05",
    doi = "10.1007/JHEP07(2011)126",
    journal = "JHEP",
    volume = "07",
    pages = "126",
    year = "2011"
}

@article{CarloniCalame:2003yt,
    author = "Carloni Calame, C. M. and Montagna, G. and Nicrosini, O. and Piccinini, F.",
    editor = "Incagli, Marco and Venanzoni, G.",
    title = "{The BABAYAGA event generator}",
    eprint = "hep-ph/0312014",
    archivePrefix = "arXiv",
    doi = "10.1016/j.nuclphysbps.2004.02.008",
    journal = "Nucl. Phys. B Proc. Suppl.",
    volume = "131",
    pages = "48--55",
    year = "2004"
}

@article{CarloniCalame:2019dom,
    author = "Carloni Calame, Carlo M. and Chiesa, Mauro and Montagna, Guido and Nicrosini, Oreste and Piccinini, Fulvio",
    title = "{Electroweak corrections to $e^+e^-\to\gamma\gamma$ as a luminosity process at FCC-ee}",
    eprint = "1906.08056",
    archivePrefix = "arXiv",
    primaryClass = "hep-ph",
    doi = "10.1016/j.physletb.2019.134976",
    journal = "Phys. Lett. B",
    volume = "798",
    pages = "134976",
    year = "2019"
}

@article{Carloni:2019ejp,
    author = "Carloni, Carlo M. and Chiesa, M. and Montagna, G. and Nicrosini, O. and Piccinini, F.",
    editor = "Blondel, A. and Gluza, J. and Jadach, S. and Janot, P. and Riemann, T.",
    title = "{$ e^+ e^-\to\gamma\gamma$ at large angle for FCC-ee luminometry}",
    doi = "10.23731/CYRM-2020-003.71",
    journal = "CERN Yellow Reports: Monographs",
    volume = "3",
    pages = "71--76",
    year = "2020"
}

@article{Engel:2025fmp,
    author = "Engel, Tim and Rocco, Marco and Signer, Adrian and Ulrich, Yannick",
    title = "{Low-energy $e^+\,e^-\to\gamma\,\gamma$ at NNLO in QED}",
    eprint = "2512.22929",
    archivePrefix = "arXiv",
    primaryClass = "hep-ph",
    month = "12",
    year = "2025"
}

@article{Budassi:2024whw,
    author = "Budassi, Ettore and Carloni Calame, Carlo M. and Ghilardi, Marco and Gurgone, Andrea and Montagna, Guido and Moretti, Mauro and Nicrosini, Oreste and Piccinini, Fulvio and Ucci, Francesco P.",
    title = "{Pion pair production in $e^+e^-$ annihilation at next-to-leading order matched to Parton Shower}",
    eprint = "2409.03469",
    archivePrefix = "arXiv",
    primaryClass = "hep-ph",
    doi = "10.1007/JHEP05(2025)196",
    journal = "JHEP",
    volume = "05",
    pages = "196",
    year = "2025"
}

@misc{Budassi:2026,
  author       = {Ettore Budassi and Carlo M. Carloni Calame and Marco Ghilardi and Andrea Gurgone and Guido Montagna and Mauro Moretti and Oreste Nicrosini and Fulvio Piccinini and Francesco P. Ucci},
  title        = {Radiative return at NLOPS accuracy},
  note         = {In preparation},
}

@misc{maestre2022precisionstudiesquantumelectrodynamics,
      title={Precision studies of quantum electrodynamics at future $e^+e^-$ colliders}, 
      author={J. Alcaraz Maestre},
      year={2022},
      eprint={2206.07564},
      archivePrefix={arXiv},
      primaryClass={hep-ph},
      url={https://arxiv.org/abs/2206.07564}, 
}

@article{PhysRevD.48.1021,
  title = {Structure function formulation of ${e}^{+}{e}^{\ensuremath{-}}\ensuremath{\rightarrow}f\overline{f}$ around the ${Z}^{0}$ resonance in a realistic setup},
  author = {Montagna, Guido and Piccinini, Fulvio and Nicrosini, Oreste},
  journal = {Phys. Rev. D},
  volume = {48},
  issue = {3},
  pages = {1021--1034},
  numpages = {0},
  year = {1993},
  month = {Aug},
  publisher = {American Physical Society},
  doi = {10.1103/PhysRevD.48.1021},
  url = {https://link.aps.org/doi/10.1103/PhysRevD.48.1021}
}

@article{CMD-3:2023alj,
    author = "Ignatov, F. V. and others",
    collaboration = "CMD-3",
    title = "{Measurement of the e+e-{\textrightarrow}{\ensuremath{\pi}}+{\ensuremath{\pi}}- cross section from threshold to 1.2~GeV with the CMD-3 detector}",
    eprint = "2302.08834",
    archivePrefix = "arXiv",
    primaryClass = "hep-ex",
    doi = "10.1103/PhysRevD.109.112002",
    journal = "Phys. Rev. D",
    volume = "109",
    number = "11",
    pages = "112002",
    year = "2024"
}

@article{Ignatov:2022iou,
    author = "Ignatov, Fedor and Lee, Roman N.",
    title = "{Charge asymmetry in e+e{\ensuremath{-}}{\textrightarrow}{\ensuremath{\pi}}+{\ensuremath{\pi}}{\ensuremath{-}} process}",
    eprint = "2204.12235",
    archivePrefix = "arXiv",
    primaryClass = "hep-ph",
    doi = "10.1016/j.physletb.2022.137283",
    journal = "Phys. Lett. B",
    volume = "833",
    pages = "137283",
    year = "2022"
}

@article{Colangelo:2022lzg,
    author = "Colangelo, Gilberto and Hoferichter, Martin and Monnard, Joachim and de Elvira, Jacobo Ruiz",
    title = "{Radiative corrections to the forward-backward asymmetry in $e^+e^-\to\pi^+\pi^-$}",
    eprint = "2207.03495",
    archivePrefix = "arXiv",
    primaryClass = "hep-ph",
    doi = "10.1007/JHEP08(2022)295",
    journal = "JHEP",
    volume = "08",
    pages = "295",
    year = "2022",
    note = "[Erratum: JHEP 09, 177 (2024), Erratum: JHEP 03, 217 (2025)]"
}

@book{Altmann:2025feg,
    author = "Altmann, J. and others",
    title = "{ECFA Higgs, electroweak, and top Factory Study}",
    eprint = "2506.15390",
    archivePrefix = "arXiv",
    primaryClass = "hep-ex",
    reportNumber = "CERN-2025-005",
    doi = "10.23731/CYRM-2025-005",
    isbn = "978-92-9083-700-8, 978-92-9083-701-5",
    series = "CERN Yellow Reports: Monographs",
    volume = "5/2025",
    month = "6",
    year = "2025"
}

@article{DGLAP,
  author       = {V.N. Gribov and L.N. Lipatov and G. Altarelli and G. Parisi and Y.L. Dokshitzer},
  title        = {Evolution of Parton Densities in QCD},
  journal      = {Sov. J. Nucl. Phys. / Nucl. Phys. B / Sov. Phys. JETP},
  volume       = {15, 126, 46},
  pages        = {298, 298, 641},
  year         = {1972, 1977, 1977},
  note         = {V.N. Gribov \& L.N. Lipatov, Sov. J. Nucl. Phys. 15 (1972) 298; G. Altarelli \& G. Parisi, Nucl. Phys. B 126 (1977) 298; Y.L. Dokshitzer, Sov. Phys. JETP 46 (1977) 641}
}

@article{FCC:2025lpp,
    author = "Benedikt, M. and others",
    collaboration = "FCC",
    title = "{Future Circular Collider Feasibility Study Report: Volume 1, Physics, Experiments, Detectors}",
    eprint = "2505.00272",
    archivePrefix = "arXiv",
    primaryClass = "hep-ex",
    reportNumber = "CERN-FCC-PHYS-2025-0002",
    month = "4",
    year = "2025"
}

@article{YFS1961,
  author       = {D.R. Yennie and S.C. Frautschi and H. Suura},
  title        = {The Infrared Divergence Phenomena and High-Energy Processes},
  journal      = {Annals of Physics},
  volume       = {13},
  pages        = {379--452},
  year         = {1961},
  doi          = {10.1016/0003-4916(61)90151-8}
}
\end{document}